\documentclass[prd, twocolumn, amsfonts, amssymb, amsmath, showkeys, nofootinbib, superscriptaddress]{revtex4-2}

\usepackage[english]{babel}
\usepackage[utf8]{inputenc}
\usepackage{amsthm}
\usepackage{mathtools}
\usepackage{physics}
\usepackage{graphicx}
\usepackage{float}
\usepackage{dcolumn}
\usepackage{bm}
\usepackage{xcolor}
\usepackage[pdftex, pdftitle={Article}, pdfauthor={Author}]{hyperref}
\usepackage{multirow}
\usepackage{hhline}
\usepackage{aas_macros}
\usepackage{enumitem}
\usepackage{orcidlink}
\newlist{todolist}{itemize}{2}
\setlist[todolist]{label=$\square$}

\newcommand{\msun}{\mathrm{\, M_\odot}}

\newcolumntype{P}[1]{>{\centering\arraybackslash}p{#1}}

\begin{document}

\preprint{APS/123-QED}

\title{Accelerating parameter estimation for parameterized tests of general relativity with gravitational-wave observations}

\author{Dhruv Kumar\orcidlink{0000-0001-8205-0404}}
\affiliation{Department of Physics, National Institute of Technology Agartala, Tripura 799046, India}
\affiliation{Institute for Gravitation and the Cosmos, Department of Physics, Pennsylvania State University, University Park, PA 16802, USA}
\affiliation{School of Physics and Astronomy, University of Glasgow, Glasgow G12 8QQ, United Kingdom}

\author{Ish Gupta\orcidlink{0000-0001-6932-8715}}
\email[Correspondence email address: ]{ishgupta@berkeley.edu}
\affiliation{Department of Physics, University of California, Berkeley, CA 94720, USA}
\affiliation{Department of Physics and Astronomy, Northwestern University, 2145 Sheridan Road, Evanston, IL 60208, USA}
\affiliation{Center for Interdisciplinary Exploration and Research in Astrophysics (CIERA), Northwestern University, 1800 Sherman Ave, Evanston, IL 60201, USA}

\author{Bangalore Sathyaprakash\orcidlink{0000-0003-3845-7586}}
\affiliation{Institute for Gravitation and the Cosmos, Department of Physics, Pennsylvania State University, University Park, PA 16802, USA}

\begin{abstract}
Tests of general relativity (GR) with gravitational waves (GWs) introduce additional deviation parameters in the waveform model. The enlarged parameter space makes inference computationally costly, which has so far limited systematic, large-scale studies that are essential to quantify parameter degeneracies, check the effect of waveform systematics, and assess robustness across non-stationary and non-Gaussian noise effects. The need is even sharper for next-generation observatories where signals are longer, signal-to-noise ratios (SNRs) are higher, and likelihood evaluations increase substantially. We address this by applying relative binning to the TIGER framework for parameterized tests of GR. Relative binning replaces dense frequency waveform evaluations with evaluations on adaptively chosen frequency bins, reducing the cost per likelihood call while preserving posterior accuracy. Using simulated binary black hole signals, we demonstrate unbiased recovery for GR-consistent cases and targeted non-GR deviations, and we map how bin resolution controls accuracy, with finer binning primarily required for the $-1$ post-Newtonian term. A high-SNR simulated signal at next-generation sensitivity further shows accurate recovery with tight posteriors. Applied to GW150914 and GW250114, both single and multi-parameter TIGER analyses finish within a day, yielding bounds consistent with GR at 90\% credibility and in agreement with previous results. Across analyses, the method reduces wall time by factors of $\mathcal{O}(10)$ to $\mathcal{O}(100)$, depending on frequency range and binning scheme, without degrading parameter estimation accuracy.

\end{abstract}

\maketitle

\section{Introduction} \label{sec:intro}

Gravitational waves (GWs) provide a unique way to test general relativity (GR), especially in the highly dynamical, strong-field regime. The detection of GW150914 by the LIGO detectors~\citep{LIGOScientific:2016aoc, Abbott2016GW150914} marked a pivotal moment in both astrophysics and GW astronomy. The analysis of GW150914 subjected the observed data to rigorous testing against GR's predictions \cite{LIGOScientific:2016lio}. 
Notably, the mass and spin of the resultant black hole inferred from the low-frequency inspiral and observed from the higher-frequency merger-ringdown signals were in excellent agreement, in accordance with GR. The event also provided stringent empirical constraints on deviations of post-Newtonian (PN) coefficients that describe the GW phase evolution. No significant deviations were observed, reinforcing GR's robustness.

Generally, parameterized tests with GW observations introduce free parameters in the GW waveform to capture violation of GR \cite{Yunes:2009ke}. One such test introduces deviations to the phase of the waveform at each PN order, allowing for systematic exploration of deviations from GR predictions \cite{Li:2011cg,Agathos:2013upa, Mehta:2022pcn}. Bayesian parameter estimation is then performed to constrain these free parameters, quantifying the inconsistency of observed GW data with GR. Posterior probability distributions of the deviation parameters consistent with zero provide evidence supporting GR, while statistically significant non-zero values could indicate the need for modified gravity theories. Compared to the double pulsar system J0737-3039~\cite{Kramer_2006}, GW150914 provided more stringent constraints on deviations in the higher PN order terms (up to 3.5PN) \cite{LIGOScientific:2016lio}. On the other hand, the first binary neutron star merger detection made by the LIGO and Virgo~\cite{VIRGO:2014yos} detectors, GW170817 \cite{LIGOScientific:2017vwq}, constrained the $-1$PN order term to $\mathcal{O}(10^{-5})$ \cite{LIGOScientific:2018dkp}, while providing constraints for other PN order terms to be comparable to that from GW150914. These bounds have been further improved by a few factors with binaries in the GWTC-3 catalog \citet{LIGOScientific:2021sio}. 

The constraints on deviations from GR will sharpen with larger samples and with high-fidelity, high signal-to-noise ratio (SNR) events~\cite{Perkins:2020tra,Gupta:2023lga}. Sensitivity upgrades to the current network, such as A+ for LIGO\cite{LIGOScientific:2007fwp,LIGOScientific:2014pky}, and proposed next-generation (XG) observatories like Cosmic Explorer\cite{evans2021horizonstudycosmicexplorer,Evans:2023euw,Gupta:2023lga} and the Einstein Telescope\cite{Punturo2010ET,Branchesi:2023mws,ET:2025xjr}, will extend the observable bandwidth and increase SNRs by more than an order of magnitude relative to present facilities. This progress exacerbates computational costs for parameterized tests. The parameter space grows once deviation coefficients are introduced, and the number of likelihood evaluations needed to resolve narrow posteriors for high-SNR signals increases accordingly. Several acceleration strategies have been proposed~\citep{Canizares:2013ywa,Smith:2016qas,Cornish:2021lje,Zackay:2018qdy,Finstad_2020,Morisaki:2021ngj,Leslie2021,Roulet_2022,islam2022factorizedparameterestimationrealtime, wong2023fastgravitationalwaveparameter, narola2023relativebinningcompletegravitationalwave,Green:2020dnx,Dax:2021tsq,Dax:2024mcn,Garcia-Quiros:2025usi,Negri:2025cyc}, but their application to GR tests has been limited~\cite{Adhikari:2022mbj}.

At current sensitivities, fast parameter estimation for GR tests enables large ensemble studies that are otherwise infeasible. Such ensembles are necessary to quantify biases from waveform modeling errors, non-stationary and non-Gaussian noise features, and missing physics, any of which can mimic apparent violations of GR~\cite{Gupta:2024gun}. For XG detectors, the longer signals and higher SNRs make full Bayesian inference with parameterized deviations prohibitively expensive without acceleration. Hence, fast and unbiased likelihood evaluation is critical to replace Fisher-matrix forecasts~\cite{Perkins:2020tra,Mahapatra:2023uwd}, which rely on linear-signal and high-SNR assumptions and tend to yield optimistic bounds~\cite{Vallisneri:2007ev,Dupletsa:2024gfl}, with full posterior constraints that faithfully capture degeneracies and non-Gaussian structure.

In this work, we show that relative binning~\cite{Zackay:2018qdy,Krishna:2023bug}, also referred to as heterodyning~\cite{Cornish:2021lje}, provides an effective route to accelerate parameter estimation with parameterized waveforms for GR tests. We implement the method within the TIGER (Test Infrastructure for General Relativity) framework~\cite{Agathos:2013upa,Li:2011cg,Roy:2025gzv}, which introduces PN deformations to the GW phase. By evaluating the waveform at adaptively chosen bin edges and interpolating within bins, the relative binning technique reduces likelihood evaluation cost while preserving posterior accuracy. We verify the linear-ratio approximation for TIGER deviation parameters, perform full Bayesian inference on simulated BBH signals at current sensitivities for GR-consistent and injected non-GR cases. We also study a high-SNR Cosmic Explorer injection and accurately recover the injected deviation with tight posteriors. Finally, utilizing public data made available by the LIGO-Virgo-KAGRA Collaboration \cite{LIGOScientific:2025snk,KAGRA:2023pio,LIGOScientific:2019lzm}, we apply the single-parameter and multi-parameter TIGER tests to GW150914 and GW250114~\cite{Abac2025}, obtaining bounds consistent with previous results. Across these tests, the method achieves $\mathcal{O}(10)$ to order $\mathcal{O}(100)$ speedups, with bin resolution chiefly impacting the $-1$PN term. Note that relative binning can also be used to accelerate other frequency-domain tests of GR (e.g., frameworks discussed in Refs.~\cite{Mehta:2022pcn,Cotesta:2020qhw,Ghosh:2016qgn,Ghosh:2017gfp,Puecher:2022sfm,Gupta:2025xxx}). In this work, we restrict ourselves to TIGER.

The paper is organized as follows. Section~\ref{sec:methods} reviews the Bayesian framework, summarizes the relative binning algorithm, and describes the TIGER pipeline. Section~\ref{sec:Benchmarking} presents validation studies on simulated BBH signals, and the high-SNR Cosmic Explorer forecast.
Section~\ref{sec:bench-real_GW_event} demonstrates the application of single and multi-parameter TIGER tests, accelerated by relative binning, to GW150914 and GW250114 and reports the results.
Section~\ref{sec:conclusion} summarizes the results and discusses implications for scalable GR tests with current and XG detectors.


\section{Methods} \label{sec:methods}
We assess relative binning as a viable acceleration technique for parameterized tests of GR by performing full Bayesian inference with TIGER phase-deformed waveforms. This section details the statistical formulation, the TIGER model, and the acceleration scheme, and specifies the diagnostics we use to validate the assumptions and quantify the computational cost. Section~\ref{subsec:methods_pe} presents the Bayesian parameter estimation setup. Section~\ref{subsec:methods_tiger} summarizes the TIGER parameterization adopted here, including the inspiral PN coefficients varied. Section~\ref{subsec:methods_rb} describes the relative binning algorithm, demonstrates the effect of binning choices, and reports measured speedups for current and XG frequency bands.

\subsection{Bayesian parameter estimation framework} \label{subsec:methods_pe}

GW parameter estimation employs Bayesian inference to extract physical parameters from detector data. The observed GW strain is modeled as:
\begin{equation}
\mathcal{D}(t) = n(t) + h(t;\theta),
\label{eq:detector_response}
\end{equation}
where $n(t)$ represents the detector noise and $h(t;\theta)$ denotes the GW signal parametrized by a vector $\theta$ containing both intrinsic parameters (component masses, spin magnitudes, and orientations) and extrinsic parameters (sky location, luminosity distance, orbital inclination, and polarization angle).

The posterior, $p(\theta|\mathcal{D})$, gives the desired probability distribution for parameters $\theta$, given the observed data $d$, and is obtained through Bayes' theorem:
\begin{equation}
p(\theta|\mathcal{D}) = \frac{\mathcal{L}(\mathcal{D}|\theta)\,\pi(\theta)}{\mathcal{Z}},
\label{eq:posterior}
\end{equation}
where $\mathcal{L}(\mathcal{D}|\theta)$ is the likelihood function, $\pi(\theta)$ encodes prior information about the parameters, and $\mathcal{Z} = \int \mathcal{L}(\mathcal{D}|\theta)\,\pi(\theta)\,\mathrm{d}\theta$ is the Bayesian evidence. For a specific parameter of interest, $\theta_0$,
the marginalized posterior distribution is obtained by integrating over all other parameters:
\begin{equation}
p(\theta_0|\mathcal{D}) = \int p(\theta|\mathcal{D}) \prod_{\theta_i \neq \theta_0} \mathrm{d}\theta_i.
\label{eq:marginalization}
\end{equation}

Computing these posterior distributions presents a substantial computational challenge, typically addressed using stochastic sampling algorithms~\cite{Gregory1992,Wood1952,hastings1970monte}. Adequate exploration of the prior volume and mapping of the likelihood surface, especially for high-SNR signals, requires $\mathcal{O}(10^8)$ likelihood evaluations. Each likelihood evaluation requires calculating the waveform, the length of which scales proportionally with the duration of the signal. Thus, the challenge is particularly acute for long-duration, high-SNR signals that will be routinely observed by XG GW observatories.

\subsection{TIGER framework} \label{subsec:methods_tiger}

\begin{figure*} 
    \centering 
    \includegraphics[width=0.9\linewidth]{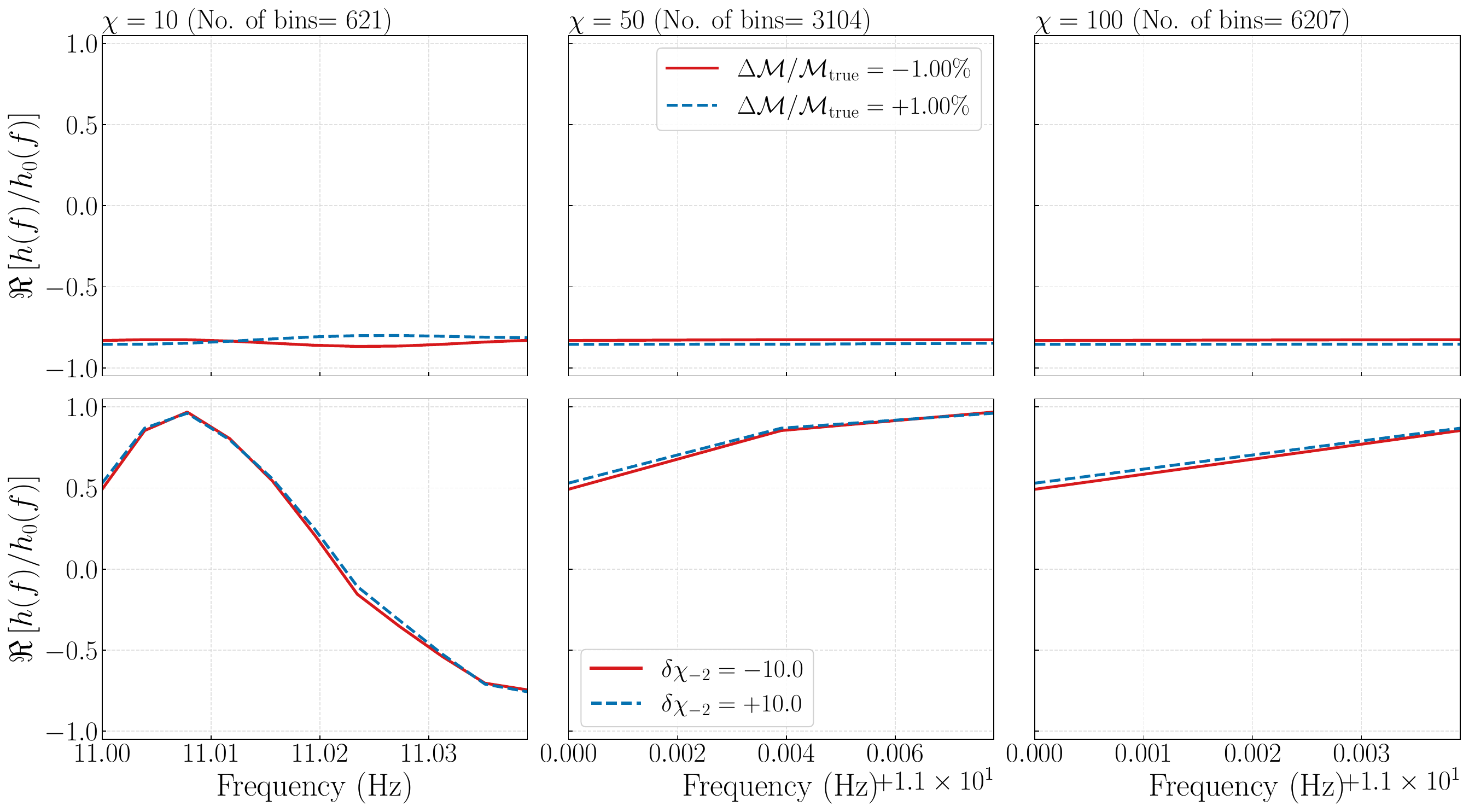} 
    \caption{
        Validation of the linear approximation in relative binning across parameter variations and binning resolutions for a BBH system with $m_1 = 25\,M_\odot$ and $m_2 = 20\,M_\odot$.
        Columns correspond to binning resolutions: $\chi = 10$ ($621$ bins; left), $\chi = 50$ ($3104$ bins; center), and $\chi = 100$ ($6207$ bins; right). 
        First row shows chirp mass variations $\Delta\mathcal{M}/\mathcal{M}_{\mathrm{true}} = \pm 1\%$; second row shows variations in the $-1$PN deviation parameter, $d\chi_{-2} = \pm 10$. 
        The real part of the waveform ratio, $\mathfrak{R} \left[ {h}(f) /h_0(f) \right]$, is plotted as a function of frequency, with blue (red) curves representing positive (negative) parameter perturbations. Variations are shown in the first frequency bin for each binning scheme. In the chosen bins, all binning schemes show fairly linear waveform ratio for $\mathcal{M}$. However, for $d\chi_{-2}$, coarse binning ($\chi=10$; left) exhibits severe non-linear artifacts and non-monotonic behavior within the bin. 
        Intermediate binning ($\chi=50$; center) shows improved linearity, and fine binning ($\chi=100$; right) achieves excellent linear approximation, validating Eq.~\eqref{eq:linear_approx}.
    } 
    \label{fig:chirp_mass_variation_comparison} 
\end{figure*}

TIGER provides a parameterized approach to testing deviations from GR by introducing controlled deformations to the GW phase. In the frequency domain, the phase can be expressed as,
\begin{equation}
\Psi(f) = 2\pi f t_c - \phi_0 - \frac{\pi}{4}
+ \sum_{j=0}^{7} \left[\psi_j + \psi^{(l)}_j \ln f \right] f^{(j-5)/3},
\label{eq:TIGER-eq1}
\end{equation}
where $t_c$ is the coalescence time and $\phi_0$ is the reference phase, and the coefficients $\psi_j$ and $\psi^{(l)}_j$ are the PN phase terms in GR, expressed as functions of the intrinsic binary parameters. TIGER changes these GR coefficients to
\begin{equation}
\psi_i = (1 + d\chi_i)\,\psi_i^{\rm GR},
\label{eq:TIGER-pg-2}
\end{equation}
so that $d\chi_i$ encodes a fractional deviation from the GR prediction at the corresponding PN order. Bayesian inference on $d\boldsymbol{\chi}$ with GW data yields bounds on deviations from GR.

Although TIGER can be extended beyond the inspiral to include merger and ringdown modifications~\cite{Roy:2025gzv}, in this work we follow the LIGO-Vrigo-KAGRA analyses~\cite{LIGOScientific:2021sio} and restrict attention to inspiral-phase deformations. We vary coefficients from the $-2$PN dipole term through 3.5PN,
\[
\{d\chi_{-2},\,d\chi_{0},\,d\chi_{1},\,d\chi_{2},\,d\chi_{3},\,d\chi_{4},\,d\chi_{5l},\,d\chi_{6},\,d\chi_{6l},\,d\chi_{7}\},
\]
where the subscript $l$ denotes the logarithmic PN contributions that first appear at 2.5PN and higher.

In the results reported in Sec.~\ref{sec:Benchmarking}, we perform Bayesian parameter estimation on simulated GR and non-GR signals while varying one deviation parameter at a time, with a uniform prior $\mathcal{U}[-10,10]$ unless stated otherwise. This one-at-a-time strategy mirrors standard practice and mitigates degeneracies among deviation parameters. Note that TIGER analyses enlarge the parameter space relative to standard inference (Sec.~\ref{subsec:methods_pe}), increasing the prior volume and the cost of sampling. Techniques involving joint variations of multiple deviation parameters (see, for e.g., Ref.~\cite{Mahapatra:2025cwk}) further worsen exploration efficiency, making acceleration techniques essential for large-scale studies.

\subsection{Relative binning framework} \label{subsec:methods_rb}

To address the computational bottleneck inherent in GW parameter estimation, several accelerated likelihood evaluation techniques have been developed~\cite{Canizares_2015,Smith_2016,Morisaki_2020}, including the relative binning method. The key insight behind relative binning is that, in frequency domain, the ratio of the waveform evaluated at two nearby points in the parameter space varies smoothly in small frequency bins. Consider a candidate waveform $h(f; \theta)$ and a fiducial reference waveform $h_0(f; \theta_0)$ chosen at the maximum likelihood point. When $\theta$ is sufficiently close to $\theta_0$, the waveform ratio
\begin{equation}
    r(f) = \frac{h(f; \theta)}{h_0(f; \theta_0)}
    \label{eq:waveform_ratio}
\end{equation}
can be approximated as linear within small frequency bins $b$:
\begin{equation}
    r(f) \approx r_0(b) + r_1(b)(f - f_m(b)),
    \label{eq:linear_approx}
\end{equation}
where $r_0(b)$ and $r_1(b)$ are linear fit coefficients determined at the bin edges, and $f_m(b)$ is the central frequency of the bin. The likelihood evaluation corresponding to $h(f; \theta)$ simplifies to evaluating the waveform only at the bin edges, significantly reducing the computational cost.

The validity of the linear approximation within each bin, and the corresponding likelihood evaluation accuracy, depends crucially on the size of the bin. In fact, it is directly related to the phase difference between the candidate waveform and the fiducial waveform within each bin. If the phase difference varies significantly across a bin, the waveform ratio exhibits oscillations, violating the linearity assumption and introducing inaccuracies in the likelihood evaluation. Conversely, if the phase difference remains small across a bin, the ratio varies smoothly and the linear approximation remains accurate. Therefore, frequency bins are constructed by controlling the accumulated phase change within a bin through a tunable parameter $\chi$\footnote{The actual implementation of relative binning has two such tunable parameters, $\epsilon$ and $\chi$~\cite{Zackay:2018qdy,Krishna:2023bug}, both of which can be adjusted to provide a tolerance on phase change across a bin. We choose to vary $\chi$ while keeping $\epsilon$ fixed.}. Larger values of $\chi$ enforce smaller phase differences between consecutive bin edges, resulting in more densely packed bins and improved accuracy of likelihood evaluation. In this way, $\chi$ provides direct control over the trade-off between computational efficiency and likelihood evaluation accuracy.

To demonstrate the dependence on bin size, we compute the waveform ratio while varying the best measured TIGER coefficient, $d\chi_{-2}$, between its prior boundaries of [-10,10], and compare it to variations in the chirp mass $\mathcal{M}$ within $1\%$ of its true value. Figure~\ref{fig:chirp_mass_variation_comparison} shows three resolutions, $\chi=10$ (621 bins), $\chi=50$ (3104 bins), and $\chi=100$ (6207 bins), with the first row varying $\mathcal{M}$ and the second row varying $d\chi_{-2}$. At $\chi=10$, the ratio is fairly linear for chosen perturbations in $\mathcal{M}$, while it displays clear oscillations for $d\chi_{-2}$, signaling a breakdown of the linear model in Eq.~\eqref{eq:linear_approx}. Increasing to $\chi=50$ suppresses these variations and yields monotonic, linear behavior for both parameters. This further improves with $\chi=100$. Thus, binning choice depends on the parameters in the waveform, the prior ranges on those parameters, the frequency range, and the SNR of the signal. We adopt $\chi = 10$ and $\chi = 50$ as practical binning parameters for subsequent analyses and benchmarking at current detector sensitivity, and using the conservative $\chi=50$ scheme when constraining deviation parameters with XG observatories.

While increasing the number of bins improves accuracy, it also raises the number of evaluations and the cost per likelihood call. To quantify this tradeoff, we measure per-call computation time on a single CPU for the exact likelihood $(T_{\rm exact})$ and for relative binning likelihood $(T_{\rm RelBin})$, then report the speedup $ \equiv T_{\rm exact}/T_{\rm RelBin}$. We use a $25\msun+20\msun$ binary in two bands that represent current and XG use cases, 20–1024 Hz (duration 16\,s) and 5–2048 Hz (duration 256\,s), respectively, and evaluate speedups for three binning configurations, $\chi\in\{10,50,100\}$. Results are summarized in Table~\ref{tab:relative_binning_speedup}.

\begin{table}[t]
\centering
\caption{Comparison between exact $(T_{\rm exact})$ and relative binning $(T_{\rm RelBin})$ likelihood calculation times for a BBH systems with masses $m_1 = 25\,M_\odot$ and $m_2 = 20\,M_\odot$. Speedup factors $(T_{\rm exact}/T_{\rm RelBin})$ are shown for two frequency bands: 20--1024\,Hz (16\,s duration) and 5--2048\,Hz (256\,s duration) across three binning parameter values: $\chi\in\{10, 50, 100\}$.}
\label{tab:relative_binning_speedup}
\begin{tabular}{|l|c|c|}
\hline
\textbf{} & \textbf{20--1024\,Hz} & \textbf{5--2048\,Hz} \\
\hline
Duration [s] & 16 & 256 \\ 
\hline
$T_{\rm exact}$ [ms] & 12.5 & 163.5  \\
\hline
\multicolumn{3}{|c|}{\textbf{$\chi=10$}} \\
\hline
$T_{\rm RelBin}$ [ms] &  0.3  & 0.4 \\
\hline
Speedup Factor & 44 & 420 \\
\hline
\multicolumn{3}{|c|}{\textbf{$\chi=50$}} \\
\hline
$T_{\rm RelBin}$ [ms] & 0.4 & 0.6 \\
\hline
Speedup Factor & 28 & 262 \\
\hline
\multicolumn{3}{|c|}{\textbf{$\chi=100$}} \\
\hline
$T_{\rm RelBin}$ [ms] & 0.6  &  0.9  \\
\hline
Speedup Factor & 20 & 163 \\
\hline
\end{tabular}
\end{table}

\begin{figure*}
    \centering
    \includegraphics[width=\linewidth]{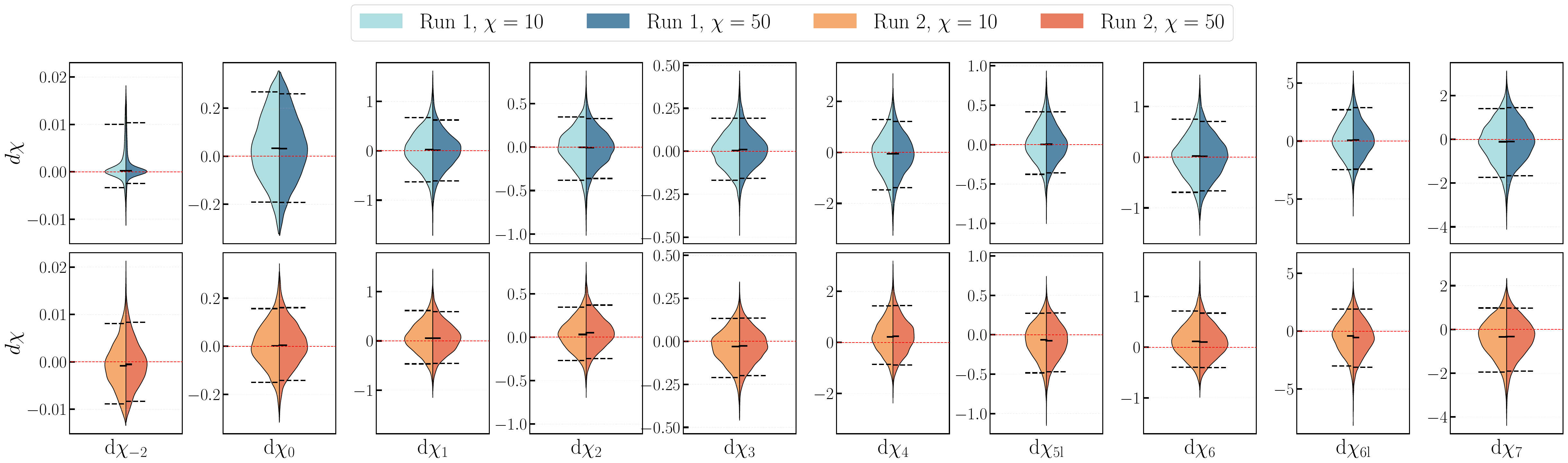}
    \caption{Posterior distributions for TIGER parameters from GR parameter estimation runs. Violin plots compare results from two independent runs (Run 1 and Run 2) at different bin numbers ($\chi = 10$ and $\chi = 50$). All posteriors are consistent and centered around zero, as expected for GR signals.}
    \label{fig:violin_plot_comparison_GR_runs.pdf}
\end{figure*}

These results show consistent $\mathcal{O}(10)$ to $\mathcal{O}(100)$ accelerations. The speedup increases with waveform duration because the exact method scales with the length of the waveform, whereas the binned evaluation scales with the number of bins. For $\chi=10$, we obtain a speedup of $\simeq 44$ in the 16\,s band and $\simeq 420$ in the 256\,s band. As $\chi$ increases, more bins are used and the advantage decreases, with speedup of $\simeq 28$–$262$ at $\chi=50$ and $\simeq 20$–$163$ at $\chi=100$. Even the conservative $\chi=100$ setting yields large gains for long signals relevant to XG sensitivities. Absolute times depend on implementation details and hardware, but the relative trends with duration and $\chi$ are robust and align with the expected scaling. Taken together, these results indicate that relative binning renders full Bayesian analyses, especially with TIGER deviation parameters, tractable for large ensembles at current sensitivities and for the long, high-SNR signals anticipated in XG detectors.

\section{Benchmarking} \label{sec:Benchmarking}

In this section, we benchmark accelerated TIGER parameter inference using full Bayesian analyses with relative binning. We analyse three sets of simulated injections in zero noise. First, we simulate GR-consistent BBH signals at the Advanced LIGO-Virgo sensitivity in the fourth observing run~\cite{ligoLIGOT2000012v2Noise,Capote:2024rmo} and show unbiased recovery of deviation parameters. Second, still at current sensitivity, we simulate systems with phase deviations and recover those deviations within the TIGER formalism. Finally, as a proof of principle, we simulate a phase-modified signal in XG sensitivity and show accurate and precise parameter recovery in less than a day for the long-duration, high-SNR signal, compared to a traditional analyses which may have taken $\sim 100$ longer.

For simulations at current sensitivities, we consider two BBH configurations: Run 1 with $(m_1,m_2)=(32\,M_\odot,8\,M_\odot)$ and Run 2 with $(25\,M_\odot,20\,M_\odot)$. Both use an effective precessing spin $\chi_p=0.2$, a luminosity distance $D_L=1000$~Mpc, and an inclination $\iota=\pi/3$. Run 1 corresponds to a system with mass-asymmetry and higher-order harmonic content~\cite{Mills:2020thr,Gupta:2024bqn}, whereas Run 2 represents a more mass-symmetric system. While a mode-by-mode relative binning technique~\cite{Leslie2021} is recommended for systems with higher harmonics, we show that at the current sensitivities, the original relative binning algorithm suffices\footnote{For usual BBH parameter estimation, \citet{Krishna:2023bug} show that the original relative binning technique produces unbiased results even for mass-asymmetric systems at current sensitivities. \citet{Gupta:2024bqn} demonstrates the same for high mass ratio systems in XG sensitivities.}, even in the context of tests of GR.

Signals are injected into the LIGO-Hanford, LIGO-Livingston, and Virgo detectors at O4 sensitivity~\cite{ligoLIGOT2000012v2Noise} in zero noise. The corresponding network SNRs are approximately 14 for Run~1 and 20 for Run~2. Waveforms are generated with the TIGER implementation of \texttt{IMRPhenomXPHM}~\cite{Pratten_2021,Roy:2025gzv} over 11–1024~Hz. Relative binning is applied with two bin resolutions, $\chi\in\{10,50\}$.

\begin{figure*}[t]
    \centering
    \makebox[\textwidth][c]{%
        \includegraphics[width=1
        \textwidth]{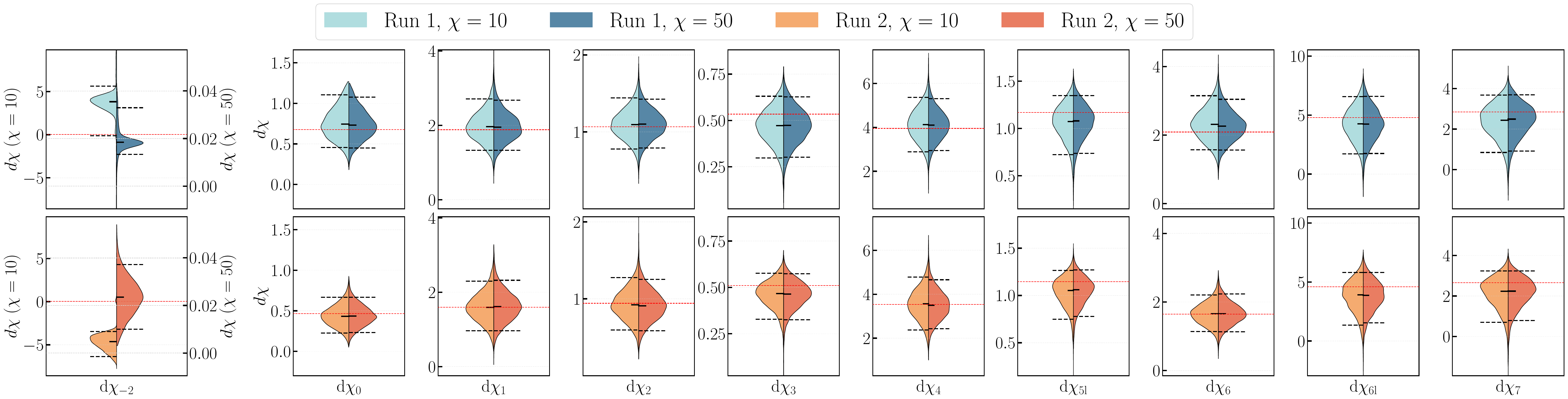}%
    }
    \caption{Posterior distributions for TIGER parameters from non-GR parameter estimation runs. Violin plots compare results from Run 1 and Run 2 at different bin numbers ($\chi = 10$ and $\chi = 50$). Most non-GR deviations are accurately recovered with the $\chi=10$ binning scheme. For $d\chi_{-2}$, using $\chi=10$ gives biased estimation, but increasing the number of bins with $\chi=50$ returns accurate posterior distributions.}
    \label{fig:Non_posterior_comparison_paper.pdf}
\end{figure*}

Parameter estimation is performed with \texttt{Bilby}~\cite{Ashton2019} and \texttt{Bilby~TGR}~\cite{2024zndo..10940210A}, using the \texttt{dynesty} sampler~\cite{Speagle:2019ivv,sergey_koposov_2025_17268284} configured with 1024 live points and 12 parallel workers. To mirror a standard TIGER analysis, we sample all intrinsic and extrinsic parameters together with a single TIGER deviation parameter at a time, adopting uniform priors $\mathcal{U}[-10,10]$ for $\{d\chi_{-2},\,d\chi_{0},\,d\chi_{1},\,d\chi_{2},\,d\chi_{3},\,d\chi_{4},\,d\chi_{5l},\,d\chi_{6},\,d\chi_{6l},\,d\chi_{7}\}$.

Section~\ref{subsec:bench-gr} validates recovery of $d\chi$ for GR-consistent injections. Section~\ref{subsec:bench-nongr} presents constraints for injected non-GR deviations constructed within TIGER. Section~\ref{subsec:bench-xg} demonstrates an XG study by injecting a non-GR signal at Cosmic Explorer sensitivity and recovering the targeted deviation with the relative binning workflow.

\subsection{GR-consistent signals} \label{subsec:bench-gr}

We first assess the performance of relative binning for GR-consistent signals using the Run~1 and Run~2 configurations. Relative binning requires a fiducial waveform $h_0(f;\theta_0)$ close to the maximum of the likelihood. Rather than fixing $\theta_0$ to the injected values, we mirror a realistic analysis: we identify the maximum-likelihood point in the full parameter space (15 GR parameters + 1 TIGER deviation) via a differential evolution algorithm~\cite{Storn:1997uea,2020SciPy-NMeth}, fix this point as $\theta_0$, precompute the summary data for $h_0$, and then run the relative binning likelihood for posterior sampling. 

Figure~\ref{fig:violin_plot_comparison_GR_runs.pdf} shows posteriors for all deviation parameters for Run~1 (top) and Run~2 (bottom). For each parameter $d\chi_i$, the left distribution corresponds to the $\chi=10$, and the right to the $\chi=50$ binning scheme. In both runs, all posteriors are consistent with the GR value $d\chi_i=0$, indicating unbiased recovery with relative binning. The $d\chi_{-2}$ parameter is best constrained, with comparatively tight posteriors also obtained for $d\chi_0$, $d\chi_2$, and $d\chi_3$. Higher-order terms and logarithmic contributions ($d\chi_4$, $d\chi_{6l}$, $d\chi_7$) are broader, reflecting their weaker imprint on the waveform for these systems. The strong tails observed in the distributions for the lower PN order terms are due to the correlation of the deviation parameter with $\mathcal{M}$, consistent with observations in previous studies~\cite{Sanger:2024axs}. The deviation parameters are also correlated with other mass and spin parameters that characterize the GW phase, which results in small offsets between the median of the posterior distribution and the injected value, even in the case of zero noise (for e.g., see Refs.~\cite{Roy:2025gzv,Mahapatra:2025cwk}).

Across all coefficients, the $\chi=10$ and $\chi=50$ results agree within sampling uncertainty, with no systematic shifts or appreciable widening at the coarser setting. Parallelized over 12 cores, the maximum sampling time across the deviation parameters and the two runs was $\sim 10$ hours ($\sim 16$ hours) for the $\chi=10$ ($\chi=50$) binning scheme. For the SNRs and binary configurations considered here, we conclude that $\chi=10$ provides sufficient accuracy while retaining a larger speedup.

\subsection{Non-GR deviation recovery} \label{subsec:bench-nongr}

\begin{table}[!htbp]
\centering
\caption{Injected parameter values for non-GR runs (Run 1 and Run 2). The values are chosen to be $n\sigma$ (where $n=3$ for $d\chi_{6l}$ and $d\chi_{7}$, and $n=5$ for all other parameters) away from the median of the respective distributions obtained in Fig.~\ref{fig:violin_plot_comparison_GR_runs.pdf}, for both $\chi=10$ and $\chi=50$ configurations.}
\label{tab:Injection_values_non_GR}
\begin{tabular}{|l|c|c|}
\hline
\textbf{Parameter} & \textbf{Run 1} & \textbf{Run 2} \\
\hline
$d\chi_{-2}$ & 0.019 & 0.025 \\
\hline
$d\chi_{0}$  & 0.676 & 0.465 \\
\hline
$d\chi_{1}$  & 1.883 & 1.596 \\
\hline
$d\chi_{2}$  & 1.066 & 0.942 \\
\hline
$d\chi_{3}$  & 0.534 & 0.510 \\
\hline
$d\chi_{4}$  & 3.949 & 3.533 \\
\hline
$d\chi_{5l}$ & 1.169 & 1.146 \\
\hline
$d\chi_{6}$  & 2.089 & 1.646 \\
\hline
$d\chi_{6l}$ & 4.783$^\dagger$ & 4.589$^\dagger$ \\
\hline
$d\chi_{7}$  & 2.842$^\dagger$ & 2.654$^\dagger$ \\
\hline
\multicolumn{3}{|l|}{$^\dagger$Injection corresponds to $3\sigma$ value.} \\
\hline
\end{tabular}
\end{table}
To test our ability to detect deviations from GR, we simulate non-GR signals by injecting non-zero values of TIGER deviation parameters, one at a time. We use the constraints obtained in Sec.~\ref{subsec:bench-gr} and choose the simulated value of $d\chi$ to be one that is $5\sigma$ away from zero. This is done for all $d\chi$ parameters except $d\chi_{6l}$ and $d\chi_{7}$, where, due to larger uncertainties for these parameters, we inject the $3\sigma$ value instead, to avoid the injection value being too close to the prior boundary. Table~\ref{tab:Injection_values_non_GR} summarizes the injected value for each deviation parameter and each run. Again, following Sec.~\ref{subsec:bench-gr}, we perform the routine of finding the maximum likelihood parameters in the full parameter space, and set that as the fiducial waveform.

Figure~\ref{fig:violin_plot_comparison_GR_runs.pdf} demonstrates successful recovery of the injected deviations, with posteriors showing significant offsets from zero, and peaking at the injected value. While most parameters show excellent consistency across runs and bin resolutions, $d\chi_{-2}$ is a notable exception: the posterior corresponding to the $\chi=10$ binning scheme is completely biased, with significant probability density away from the true value, while $\chi = 50$ provides accurate and precise constraints consistent with the simulated value. This indicates that coarser binning inadequately resolves the low-frequency features crucial for measuring $d\chi_{-2}$. While the loss of linearity in waveform ratio for such a case was already indicated in Fig.~\ref{fig:chirp_mass_variation_comparison}, in Appendix~\ref{app:dch_minus2}, we also compare the erroneous likelihood evaluations in the $\chi=10$ case for the $d\chi_{-2}$ parameter, with the $\chi=50$ case where we obtain unbiased recovery.

Nevertheless, the strong agreement for remaining parameters confirms the robustness of relative binning, even with the coarser $\chi=10$ binning scheme for the considered systems. In fact, similar sampling times to those reported in Sec.~\ref{subsec:bench-gr} are also obtained for runs presented in this section. However, we caution that for the best measured parameters like $d\chi_{-2}$, a finer binning scheme like $\chi=50$, or a more stringent prior, as is tuned for $\mathcal{M}$\footnote{Relative binning is observed to perform well when $\mathcal{M}$ is restricted to a small region around the maximum likelihood value~\cite{Krishna:2023bug}.}, works better.

\subsection{Non-GR deviation recovery in XG sensitivity} \label{subsec:bench-xg}

Relative binning yields its largest gains for long-duration signals, which are characteristic of XG observatories (see Table~\ref{tab:relative_binning_speedup}). In this section, we test the method in the XG regime.

As a proof of principle, we simulate a Run~1–like BBH in Cosmic Explorer sensitivity with an injected deviation $d\chi_{3}=-0.1$, a value that lies within the 90\% credible bounds reported in GWTC-3~\cite{LIGOScientific:2021sio}. The analysis uses a 5–2048 Hz band and simulates the signal in zero noise. The resulting SNR in a 40 km Cosmic Explorer interferometer is $\sim 450$. We adopt a conservative binning choice $\chi=50$ and perform parameter estimation with the TIGER pipeline, varying all intrinsic and extrinsic parameters together with a single deviation coefficient and placing a uniform prior $\mathcal{U}[-5,5]$ on $d\chi_{3}$. The 16-dimensional parameter space is sampled in parallel over 12 cores.

\begin{figure}[!htbp]
    \centering
    \includegraphics[width=0.9\linewidth]{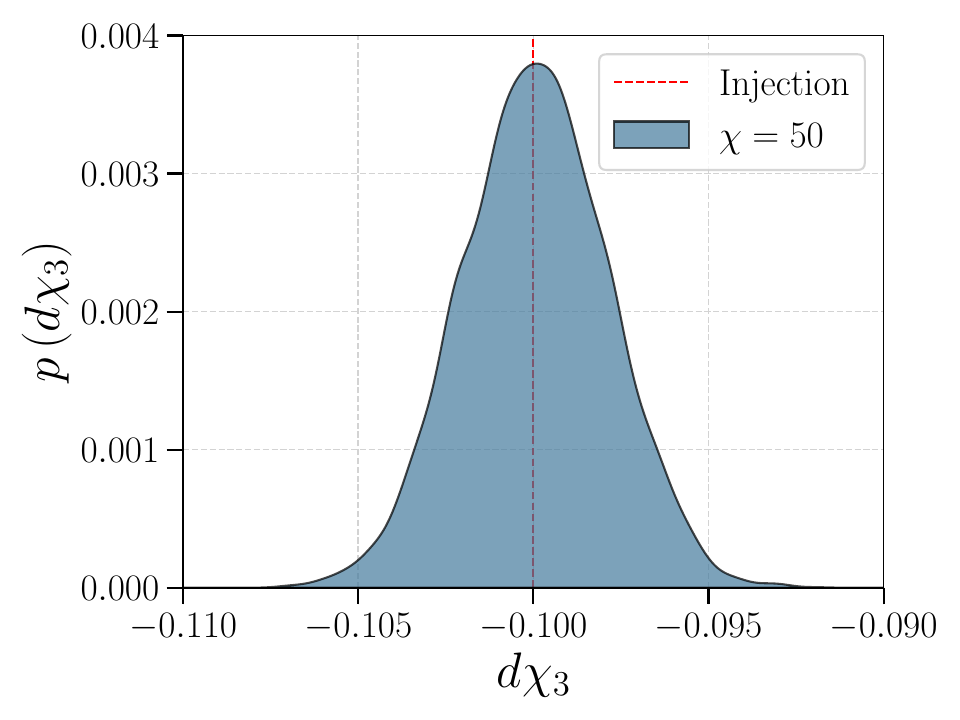}
    \caption{Posterior probability distribution showing accurate and precise inference of the parameter $d\chi_3$ for a non-GR Run 1-like simulated BBH signal in the 40 km Cosmic Explorer sensitivity. The red-dashed line shows the simulated value of $d\chi_3=-0.1$.}
    \label{fig:CE_dchi3_posterior_plot.pdf}
\end{figure}

Figure~\ref{fig:CE_dchi3_posterior_plot.pdf} shows the posterior for $d\chi_{3}$, yielding $d\chi_{3} = -0.100^{+0.002}_{-0.002}$,
which demonstrates unbiased recovery and a width consistent with the high SNR of the signal. Parallelized over 12 cores, the sampling took less than a day to finish. This illustrates that relative binning enables full Bayesian TIGER inference for XG-like durations while retaining accuracy. It also highlights the constraining power of XG observatories: for a deviation consistent with current bounds, the GR value $d\chi_{3}=0$ is excluded at better than $5\sigma$, providing a concrete example of how XG data can deliver decisive constraints on phase deformations.

\section{Application to GW150914 and GW250114} \label{sec:bench-real_GW_event}

\begin{figure*}
    \centering    \includegraphics[width=\linewidth]{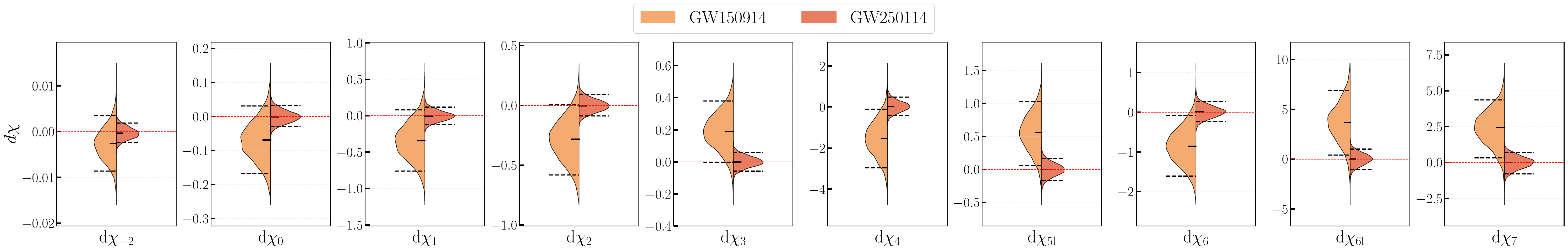}
    \caption{Posterior distributions for deviation parameters from analysis of GW150914 (left) and GW250114 (right), with the $\chi=50$ binning scheme. All posteriors are consistent with GR, with distributions significantly more constrained for GW250114 due to higher SNR. The red dashed horizontal line shows the GR value of zero. The medians and 90\% credible intervals are explicitly stated in Table~\ref{tab:GR_constraints_3dec} in Appendix~\ref{app:GW_events}.}
    \label{fig:GW_event.pdf}
\end{figure*}

We apply the TIGER pipeline with relative binning to two GW events to test performance on observed data: GW150914~\cite{LIGOScientific:2016aoc,LIGOScientific:2016lio} (SNR $\sim 24$) and the high-fidelity event GW250114~\cite{Abac2025,LIGOScientific:2025obp} (SNR $\sim 76$). The latter has enabled stringent tests of GR, including a validation of Hawking's area law~\cite{Hawking:1971tu,Abac2025} and tight constraints from multiple formalisms~\cite{LIGOScientific:2025obp}.

We conduct two complementary analyses for each event. First, we perform the standard single-parameter TIGER analysis used in the rest of this work, varying one $d\chi_i$ at a time while sampling all intrinsic and extrinsic parameters. This mapping is straightforward to interpret in terms of specific PN deformations and connections to alternative-theory coefficients~\cite{Yunes:2009ke}. Second, following Refs.~\cite{Saleem:2021nsb,Datta:2022izc,Mahapatra:2025cwk}, we perform a multi-parameter TIGER analysis in which we vary the set $\{d\chi_3,\,d\chi_4,\,d\chi_{5l},\,d\chi_6,\,d\chi_{6l},\,d\chi_7\}$ within a uniform prior of $\mathcal{U}[-20,20]$, then compute the variance–covariance matrix of these parameters and carry out a principal component analysis (PCA). The multi-parameter approach captures possible correlated deviations across PN orders in a single run, and PCA identifies the best-measured linear combinations that are most tightly constrained. To test GR, we check the consistency of the posterior distributions of these linear combinations (usually, just the first two, best constrained PCA parameters~\cite{Saleem:2021nsb,Mahapatra:2025cwk}) with the GR value of zero.

\begin{figure}
    \centering    \includegraphics[width=\linewidth]{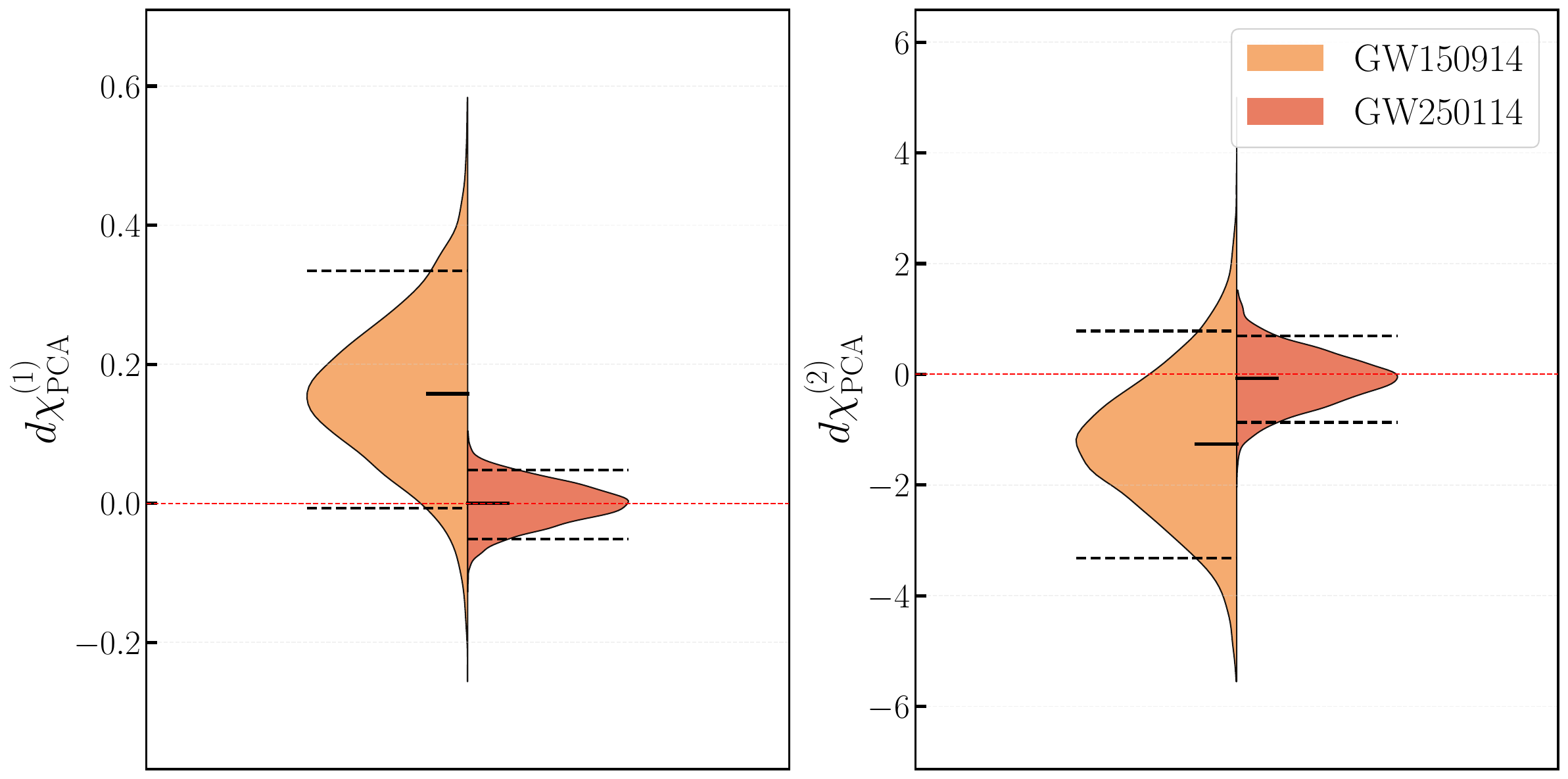}
    \caption{Constraints on the first two PCA parameters when the multi-parameter TIGER pipeline is applied to GW150914 (light orange) and GW250114 (dark orange). The horizontal red dashed line corresponds to the GR value of zero. The medians and 90\% credible intervals are explicitly stated in Table~\ref{tab:GR_constraints_3dec} in Appendix~\ref{app:GW_events}.}
    \label{fig:GW_event_PCA}
\end{figure}

The single-parameter analysis increases dimensionality only mildly and yields easily interpretable constraints. The multi-parameter analysis enlarges the parameter space by six additional parameters, increasing sampling cost substantially. Relative binning reduces the per-call likelihood cost sufficiently to make the multi-parameter runs tractable for both events, enabling a direct comparison between single-parameter bounds and PCA-constrained combinations on real data.

Figure~\ref{fig:GW_event.pdf} shows posteriors from the single-parameter TIGER analyses of GW150914 and GW250114 using the $\chi=50$ binning. For both events, most deviation parameters are consistent with the GR value of zero at 90\% credibility and agree with previous LIGO–Virgo–KAGRA results~\cite{LIGOScientific:2016lio,LIGOScientific:2025obp}. Constraints are significantly tighter for GW250114, reflecting its higher SNR. For GW150914, several posteriors place the GR value near the boundary of the 90\% credible interval, and for the subset of the 2PN and higher terms, the GR value lies just outside the interval. This qualitative behavior was also reported in Ref.~\cite{LIGOScientific:2016lio} and is consistent with parameter correlations and the effects of Gaussian noise. In contrast, for GW250114, the GR value lies close to the median for all deviation parameters.

We next consider multi-parameter variations, followed by PCA, and report the first two, best-constrained PCA combinations in Figure~\ref{fig:GW_event_PCA}. Light orange corresponds to GW150914 and dark orange to GW250114. GW250114 yields $d\chi_{\rm PCA}^{(1)} = 0.00_{-0.05}^{+0.05}$ and
$d\chi_{\rm PCA}^{(2)} = -0.08_{-0.79}^{+0.77}$, which are statistically consistent with those in Ref.~\cite{LIGOScientific:2025obp}. The trends mirror the single-parameter results: broader posteriors for GW150914 with the GR value near the 90\% credible boundary, and markedly narrower, GR-consistent posteriors for GW250114. A similar contrast between lower and higher SNR regimes for PCA combinations was observed in simulated runs in Ref.~\cite{Mahapatra:2025cwk}. Notably, the 21-dimensional multi-parameter runs complete on comparable timescales to the single-parameter analyses when using relative binning: with $\chi=50$ and 12-core parallelization, wall times are about 12 hours for GW150914 and about one day for GW250114.

These applications demonstrate that the TIGER pipeline accelerated with relative binning reproduces published constraints obtained with full likelihood evaluations while significantly reducing computation time. The gains are most pronounced for multi-parameter runs, where high-SNR events can be analyzed within a day, enabling routine PCA-based tests on observational data.

\section{Conclusions} \label{sec:conclusion} 

We have implemented and validated relative binning within the TIGER framework and shown that it enables fast and accurate implementation of parameterized tests of GR with GWs. By replacing dense frequency evaluations with evaluations of the waveform on adaptively chosen bins, the cost per likelihood call is reduced while posterior accuracy is preserved. 

We quantified the validity of the linear ratio approximation that underpins relative binning by varying a well-measured binary parameter (the chirp mass) and a well-measured TIGER deviation parameter ($d\chi_{-2}$) about a fiducial point. For coarse binning, $\chi=10$ ($\sim 600$ bins), the waveform ratio when varying chirp mass remains approximately linear, but the $d\chi_{-2}$ case exhibits significant nonlinearities (c.f. Fig.~\ref{fig:chirp_mass_variation_comparison}). Increasing resolution to $\chi=50$ ($\sim 3000$ bins) restores linearity for $d\chi_{-2}$, and $\chi=100$ ($\sim 6000$ bins) yields excellent agreement with the linear assumption. Measured speedups with relative binning range from $\mathcal{O}(10)$ to $\mathcal{O}(100)$, depending on duration and binning (c.f. Tab.~\ref{tab:relative_binning_speedup}). For 20--1024 Hz, 16\,s signals, we obtain speedup factors of 44, 28, and 20 for $\chi=10,50,100$. For 5–2048 Hz, 256\,s signals, the factors increase to about 420, 262, and 163. Thus, the method becomes increasingly advantageous for long signals that will be common in XG observatories. 

We performed a comprehensive injection campaign to establish the reliability of the acceleration technique. For GR-consistent simulated signals, parameter estimation yields posteriors centered on zero for all TIGER coefficients, with no statistical difference in recoveries between the $\chi=10$ and $\chi=50$ binning schemes (c.f. Fig.~\ref{fig:violin_plot_comparison_GR_runs.pdf}). Accelerated parameter estimation with one TIGER coefficient at a time completes in about 10 hours for $\chi=10$ and about 16 hours for $\chi=50$ on 12 cores. For non-GR injections, posteriors peak at the injected values across coefficients on similar computational timescales (c.f. Fig.~\ref{fig:Non_posterior_comparison_paper.pdf}). However, we identified a parameter-specific caveat: with the $\chi=10$ binning scheme, $d\chi_{-2}$ posterior shows biased recovery, while $\chi=50$ recovers it accurately. This suggests a practical guideline to use $\chi=50$ when constraining the $-1$PN term, at current detector sensitivities, and to reserve $\chi=100$ for very conservative analyses. 

To probe the XG regime, we analyzed a simulated BBH in Cosmic Explorer sensitivity with an injected deviation $d\chi_{3}=-0.1$ over 5–2048 Hz and SNR about 450. With $\chi=50$, the posterior yields $d\chi_{3}=-0.100^{+0.002}_{-0.002}$ in less than a day of computation (c.f. Fig.~\ref{fig:CE_dchi3_posterior_plot.pdf}). This demonstrates that full Bayesian TIGER inference, accelerated with relative binning, is feasible for XG-like duration signals that are otherwise intractable with traditional parameter inference techniques. 

We also applied the pipeline to GW150914 and GW250114 to validate the approach on observed data. Single-parameter TIGER runs reproduce published bounds at 90\% credibility and are consistent with previous LIGO–Virgo–KAGRA results (c.f. Fig.~\ref{fig:GW_event.pdf} and Table~\ref{tab:GR_constraints_3dec}). In addition, we performed multi-parameter runs varying $\{d\chi_3, d\chi_4, d\chi_{5l}, d\chi_6, d\chi_{6l}, d\chi_7\}$ jointly and conducted principal component analysis with resulting posteriors. The leading principal components are tightly constrained and statistically consistent with published results (c.f. Fig.~\ref{fig:GW_event_PCA} and Table~\ref{tab:GR_constraints_3dec}). With $\chi=50$, the 21-dimensional analyses completed in about 12 hours for GW150914 and about one day for GW250114, comparable to the single-parameter TIGER runs. These applications confirmed that the method reproduces TIGER constraints while significantly reducing computation time, with the gains most pronounced for multi-parameter runs that would otherwise be computationally prohibitive for routine application to the growing observed catalog. 

The approach is broadly applicable to other frequency-domain parameterized tests of GR. Each framework should be validated, as done here, by checking linearity in the relevant deformation directions and by checking accuracy as a function of binning. With the reduced computational burden, large-scale studies become feasible, enabling systematic studies of waveform modeling errors, non-stationary and non-Gaussian noise, calibration uncertainties, and missing physics such as eccentricity or environmental effects~\cite{Narayan:2023vhm,Gupta:2024gun,Gupta:2025paz}. While we have presented one high-SNR injection in XG sensitivity, this analysis can be extended to generate comprehensive forecasts across multiple detector configurations, including XG ground-based facilities such as Cosmic Explorer and Einstein Telescope. Moreover, with significant reduction in computational load for multi-parameter tests, this work motivates an exhaustive comparison of multi-parameter TIGER tests of general relativity with and without relative binning, especially in the context of obtaining precise estimates with principal component analyses. Taken together, these results establish relative binning as a reliable and efficient tool for parameterized tests of GR with current detectors and a necessary component for routine, large-scale, multi-parameter studies in the XG era.

\begin{acknowledgments}
The authors would like to Koustav Chandra and Aditya Vijaykumar for useful comments on the draft, and Parthapratim Mahapatra for facilitating comparisons with previous PCA analyses. DK and BS would like to acknowledge the support of the National Science Foundation via NSF grants: PHY-2207638, AST-2307147, PHY-2308886, PHY-2309064. IG acknowledges support from the Network for Neutrinos,
Nuclear Astrophysics, and Symmetries (N3AS) Collaboration, NSF grant: PHY-2020275. The authors would also like to acknowledge the LIGO Laboratory computing resources supported by NSF grants:PHY-0757058 and PHY-0823459, and the Gwave cluster maintained by the Institute for Computational and Data Sciences at Penn State University, supported by NSF grants: OAC-2346596, OAC-2201445, OAC- 2103662, OAC-2018299, and PHY-2110594. This research has made use of data or software obtained from the Gravitational Wave Open Science Center (gwosc.org), a service of the LIGO Scientific Collaboration, the Virgo Collaboration, and KAGRA. This material is based upon work supported by NSF's LIGO Laboratory which is a major facility fully funded by the National Science Foundation, as well as the Science and Technology Facilities Council (STFC) of the United Kingdom, the Max-Planck-Society (MPS), and the State of Niedersachsen/Germany for support of the construction of Advanced LIGO and construction and operation of the GEO600 detector. Additional support for Advanced LIGO was provided by the Australian Research Council. Virgo is funded, through the European Gravitational Observatory (EGO), by the French Centre National de Recherche Scientifique (CNRS), the Italian Istituto Nazionale di Fisica Nucleare (INFN) and the Dutch Nikhef, with contributions by institutions from Belgium, Germany, Greece, Hungary, Ireland, Japan, Monaco, Poland, Portugal, Spain. KAGRA is supported by Ministry of Education, Culture, Sports, Science and Technology (MEXT), Japan Society for the Promotion of Science (JSPS) in Japan; National Research Foundation (NRF) and Ministry of Science and ICT (MSIT) in Korea; Academia Sinica (AS) and National Science and Technology Council (NSTC) in Taiwan.

This work has extensively utilized the following software packages: \texttt{numpy}~\cite{harris2020array}, \texttt{scipy}~\cite{2020SciPy-NMeth}, \texttt{matplotlib}~\cite{Hunter:2007}, \texttt{bilby}~\cite{Ashton2019}, \texttt{bilby tgr}~\cite{2024zndo..10940210A}, and \texttt{dynesty}~\cite{Speagle:2019ivv,sergey_koposov_2025_17268284}.
\end{acknowledgments}

\appendix

\section{Likelihood evaluation inaccuracy for $d\chi_{-2}$}\label{app:dch_minus2}
\begin{figure}[!htbp]
    \centering
    \includegraphics[width=\linewidth]{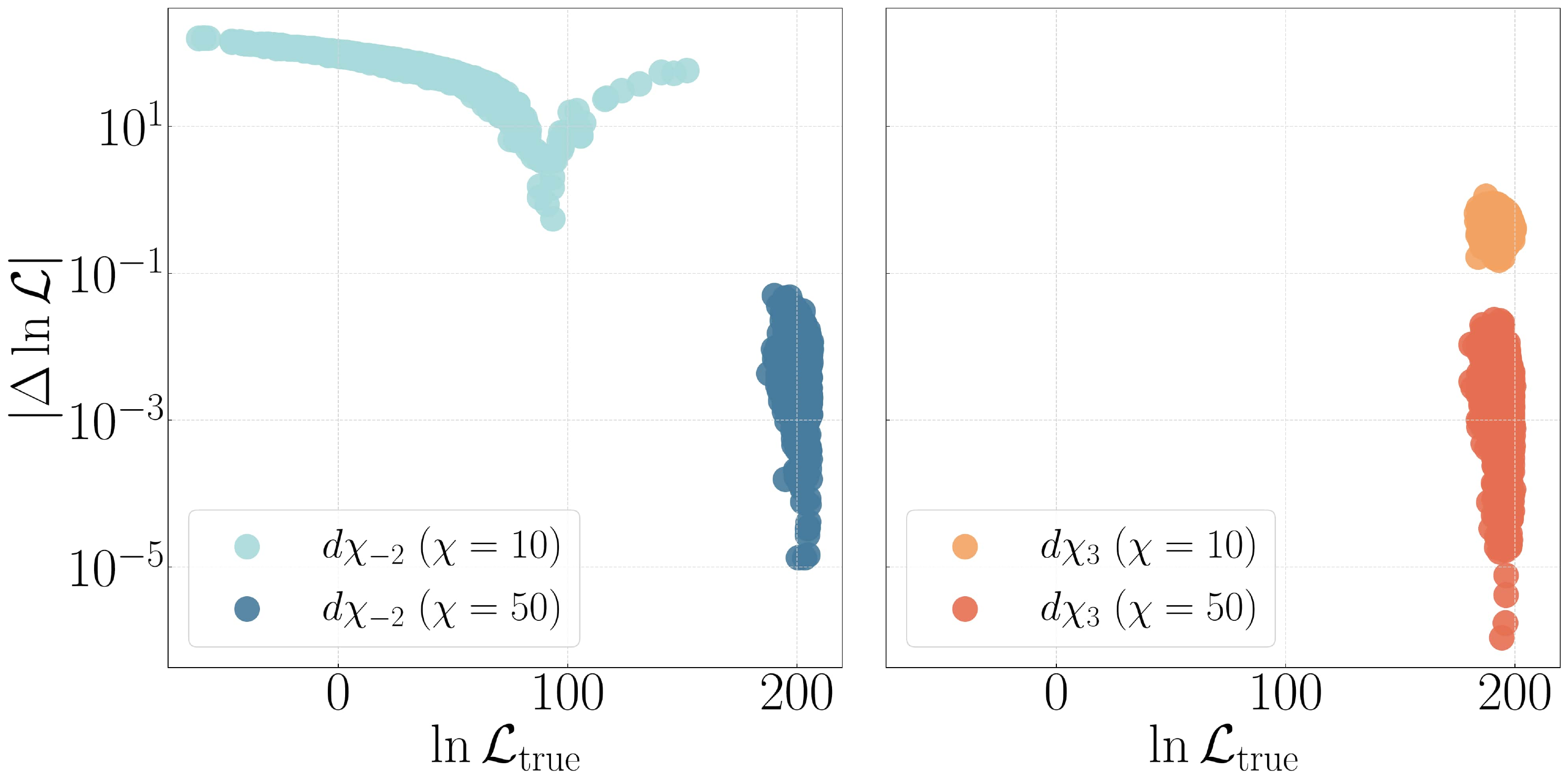}
    \caption{The absolute value of the difference between the true likelihood and the likelihood calculated using the relative binning technique, for posterior samples corresponding to Run 1 results shown in Fig.~\ref{fig:Non_posterior_comparison_paper.pdf}. The left panel shows the difference for the $d\chi_{-2}$ case, where we see biased inference for the $\chi=10$ scheme, and the right panel shows the difference for the $d\chi_{3}$ case, which corresponds to unbiased inference.}
    \label{fig:weights_analysis_plot.pdf}
\end{figure}

In Sec.~\ref{subsec:bench-nongr}, we observed biased parameter recovery for $d\chi_{-2}$ when using $\chi=10$ binning. This was resolved when a finer binning scheme, $\chi=50$, was employed. To investigate this, we compare the likelihood evaluated using the relative binning scheme with the full GW likelihood for the posterior samples for the Run 1-$d\chi_{-2}$ case. The absolute value of this difference is shown in the left panel of Fig.~\ref{fig:weights_analysis_plot.pdf}.

For the $\chi=10$ run (light blue points), we see significant difference $(\mathcal{O}(100))$ between the true likelihood and the relative binning likelihood. This is related to the significant non-linearity in the waveform ratio obtained for the $d\chi_{-2}$ case in Fig.~\ref{fig:chirp_mass_variation_comparison}. With the $\chi=50$ binning scheme, the absolute difference in likelihoods drops below $10^{-1}$, resulting in an unbiased recovery of $d\chi_{-2}$ in Fig.~\ref{fig:Non_posterior_comparison_paper.pdf}. For comparison, we also show the likelihood comparison for the $d\chi_{3}$ case (right panel in Fig.~\ref{fig:weights_analysis_plot.pdf}), where we obtain unbiased recovery for both $\chi=10$ and $\chi=50$ binning. This is consistent with the small absolute difference between the two likelihoods: less than $1$ for the $\chi=10$ case, and less than $0.1$ for the $\chi=50$ case.

\section{Constraints from GW events}
\label{app:GW_events}
In Table~\ref{tab:GR_constraints_3dec}, we present the median and 90\% credible intervals corresponding to the constraints shown in Fig.~\ref{fig:GW_event.pdf} and Fig.~\ref{fig:GW_event_PCA}. 
\begin{table}[!htbp]
\renewcommand{\arraystretch}{1.5}
\centering
\caption{Constraints on TIGER parameters from the single-parameter runs, and on the two best measured PCA parameters from the multi-parameter runs, for GW150914 and GW250114, with $\chi=10$ and $\chi=50$.}
\vspace{1em}
\label{tab:GR_constraints_3dec}
\begin{tabular}{|l|c|c|c|c|}
\hline
\textbf{Par} & \multicolumn{2}{c|}{\textbf{GW150914}} & \multicolumn{2}{c|}{\textbf{GW250114}} \\
\hline
 & $\chi=10$ & $\chi=50$ & $\chi=10$ & $\chi=50$ \\
\hline
$d\chi_{-2}$ & $-0.003_{-0.006}^{+0.006}$ & $-0.003_{-0.006}^{+0.006}$ & $-0.001_{-0.002}^{+0.003}$ & $-0.000_{-0.002}^{+0.002}$ \\
\hline
$d\chi_{0}$  & $-0.071_{-0.099}^{+0.101}$ & $-0.069_{-0.098}^{+0.100}$ & $-0.003_{-0.029}^{+0.034}$ & $-0.001_{-0.029}^{+0.033}$ \\
\hline
$d\chi_{1}$  & $-0.347_{-0.407}^{+0.406}$ & $-0.343_{-0.418}^{+0.424}$ & $-0.005_{-0.120}^{+0.138}$ & $-0.005_{-0.114}^{+0.122}$ \\
\hline
$d\chi_{2}$  & $-0.281_{-0.315}^{+0.294}$ & $-0.284_{-0.299}^{+0.292}$ & $-0.001_{-0.086}^{+0.091}$ & $-0.004_{-0.085}^{+0.093}$ \\
\hline
$d\chi_{3}$  & $0.197_{-0.185}^{+0.193}$ & $0.191_{-0.194}^{+0.189}$ & $0.001_{-0.058}^{+0.055}$ & $0.001_{-0.058}^{+0.057}$ \\
\hline
$d\chi_{4}$  & $-1.470_{-1.491}^{+1.377}$ & $-1.536_{-1.426}^{+1.434}$ & $-0.041_{-0.396}^{+0.383}$ & $0.021_{-0.438}^{+0.460}$ \\
\hline
$d\chi_{5l}$ & $0.560_{-0.472}^{+0.476}$ & $0.558_{-0.494}^{+0.476}$ & $0.001_{-0.167}^{+0.154}$ & $-0.003_{-0.164}^{+0.167}$ \\
\hline
$d\chi_{6}$  & $-0.862_{-0.739}^{+0.763}$ & $-0.855_{-0.754}^{+0.767}$ & $-0.016_{-0.221}^{+0.228}$ & $0.009_{-0.247}^{+0.253}$ \\
\hline
$d\chi_{6l}$ & $3.60_{-3.33}^{+3.12}$ & $3.68_{-3.27}^{+3.23}$ & $-0.018_{-1.026}^{+0.956}$ & $0.014_{-1.039}^{+0.990}$ \\
\hline
$d\chi_{7}$  & $2.44_{-2.04}^{+1.99}$ & $2.43_{-2.10}^{+1.93}$ & $-0.086_{-0.751}^{+0.727}$ & $0.010_{-0.793}^{+0.715}$ \\
\hline
PCA1 & $0.157_{-0.163}^{+0.175}$ & $0.158_{-0.165}^{+0.177}$ & $-0.003_{-0.055}^{+0.048}$ & $0.000_{-0.052}^{+0.048}$ \\
\hline
PCA2 & $-1.16_{-2.06}^{+2.20}$ & $-1.26_{-2.07}^{+2.04}$ & $0.120_{-0.814}^{+0.830}$ & $-0.076_{-0.793}^{+0.770}$ \\
\hline
\end{tabular}
\end{table}

\bibliography{bibliography}

\end{document}